\documentclass[varenna]{cimento}
\usepackage{comment} 
\usepackage{amsmath,amssymb}
\usepackage{lineno}
\usepackage{graphicx}
\pdfoutput=1 

\title{Two-particle angular correlations of identified particles in pp collisions at $\sqrt{s}$ = 13 TeV with ALICE}
\shorttitle{Two-particle angular correlations of identified particles in pp collisions with ALICE}
\author{Daniela Ruggiano (for the ALICE Collaboration)
}
\institute{
Faculty of Physics, Warsaw University of Technology\\ Koszykowa 75, 00-662, Poland}

\begin{document}
\maketitle
\begin{abstract}
Two-particle angular correlation is one of the most powerful tools to study the mechanism of particle production in proton--proton (pp) collision systems by relating the difference between the azimuthal angle ($\Delta\varphi$) and the rapidity ($\Delta y$) of the particles from a pair. 
Hadronization processes are influenced by various physical phenomena, such as resonance decay, Coulomb interactions, laws of conservation of energy and momentum, and others, because of the quark content of the particles involved. Therefore, each correlation function is unique and shows a different dependence on transverse momentum $p_{\mathrm{T}}$ and/or multiplicity. The angular correlation functions reported by the ALICE Collaboration in pp collisions showed an anticorrelation in short range of ($\Delta y,\Delta\varphi$) for baryon pairs which is not predicted by any theoretical model.\\
\indent
In this contribution, this behavior will be investigated by studying identified charged hadrons (i.e., $\pi^{\pm}$, $\rm K^{\pm}$, and p($\bar{\rm p}$)) in the $\Delta y,\Delta\varphi$ space in pp collisions at $\sqrt{s} = 13$ TeV recorded by ALICE. In addition, to distinguish the various physical contributions, collisions with different multiplicities are analyzed separately and diverse normalization methods are applied.
\end{abstract}
\vspace{-0.5cm}
\section{Introduction}
Ultrarelativistic proton--proton (pp) collisions at the Large Hadron Collider (LHC) offer the opportunity to study the mechanism of particle production with high precision. To understand the processes involved, several tools can be used. This work focuses on angular correlations, which offer rich information to comprehend the behavior of elementary particles by analyzing the correlations between two particles measured as a function of the relative pseudorapidity $\Delta\eta$ = $\eta_{\mathrm{1}} - \eta_{\mathrm{2}}$ and azimuthal angle $\Delta\varphi$ = $\varphi_{\mathrm{1}} - \varphi_{\mathrm{2}}$, where '1' and '2' denote the two particles. These correlations provide a tool to explore multiple mechanisms simultaneously, including (mini)jets, Bose--Einstein or Fermi--Dirac quantum statistics (in the case of identical bosons or fermions), resonance decays, conservation laws, and other sources of additional correlations, each resulting in a unique distribution in the $\Delta y,\Delta\varphi$ space.\\
\indent
Previous measurement performed in pp collisions at $\sqrt{s}$ = 7 TeV has demonstrated that there are  differences in the shapes of the global correlations for different types of particles \cite{Adam_2017}.  In particular, the baryon--baryon pairs, when combined with the antibaryon--antibaryon correlations, show a dip around ($\Delta\eta$,$\Delta\varphi$) $\sim$ (0,0), which is qualitatively different from the correlations between two mesons and the baryon--antibaryon pairs. Such behavior is not reproduced by theoretical models. To investigate this, various studies have been done focusing on different transverse momentum $p_{\mathrm{T}}$ intervals, diverse baryons, as well as the influence of the Coulomb interaction and Fermi--Dirac quantum statistics. These studies produce consistent findings, indicating the presence of anticorrelation between baryon--baryon and antibaryon--antibaryon pairs.\\
\indent
A further study involves the behaviour of the correlation functions for different multiplicity classes, as has been done in this work. The previous analysis was performed for pp collisions at $\sqrt{s}$ = 7 TeV, but interestingly, the results showed that the correlation functions decreased as multiplicity increased following a so-called trivial scaling trend of 1/N. 
To understand this phenomenon, a comprehensive analysis in different multiplicity classes by studying combinations of identified charged particles (i.e., $\rm{\pi^{\pm}}$, $\rm{K^{\pm}}$, and p($\rm{\bar{p}}$)) in the $\Delta y,\Delta\varphi$ space is performed in pp collision at $\sqrt{s}$ = 13 TeV using the probability ratio \cite{probability} and the rescaled two-particle cumulant \cite{rescaled}.

\section{Definition of two-particle correlation function}
{\textbf{Probability ratio definition}} -- The experimental correlation function is defined as
\begin{equation}
    C_{\mathrm{P}}(\Delta y,\Delta\varphi)=\frac{S(\Delta y,\Delta\varphi)}{B(\Delta y,\Delta\varphi)}.
    \label{equation:probratio}
\end{equation}
The signal distribution $S(\Delta y,\Delta\varphi)$  is constructed from particle pairs coming from the same event (collision), which can be expressed as 
\begin{equation}
    S(\Delta y,\Delta\varphi)=\frac{1}{N^{\mathrm{signal}}_{\mathrm{pairs}}}\frac{\mathrm{d^{2}}N^{\mathrm{signal}}_{\mathrm{pairs}}}{{\mathrm{d}\Delta y}\mathrm{d}{\Delta\varphi}},
\end{equation}
where $N^{\mathrm{signal}}_{\mathrm{pairs}}$ is amount of pairs of particles in signal. The background distribution $B(\Delta y,\Delta\varphi)$ is constructed from pairs of particles from different events (meaning there should be no correlation between them) and defined as 
\begin{equation}
     B(\Delta y,\Delta\varphi)=\frac{1}{N^{\mathrm{mixed}}_{\mathrm{pairs}}}\frac{\mathrm{d^{2}}N^{\mathrm{mixed}}_{\mathrm{pairs}}}{{\mathrm{d}\Delta y}\mathrm{d}{\Delta\varphi}},
\end{equation}
where $N^{\mathrm{mixed}}_{\mathrm{pairs}}$ is amont of pairs of particles in background. Each of the $S(\Delta y,\Delta\varphi)$ and $B(\Delta y,\Delta\varphi)$ distributions are divided by respective number of pairs.\\

\textbf{Rescaled two-particle cumulant definition} -- The rescaled two-particle cumulant, which was first introduced by the STAR Collaboration \cite{rescaled}, is defined as
\begin{equation}
    C_{\mathrm{C}}(\Delta y,\Delta\varphi) = \frac{N_{\mathrm{av}}}{\Delta y,\Delta\varphi} (C_{\mathrm{P}}-1),
    \label{equa:densitycorr}
\end{equation}
where $\Delta y$ and $\Delta\varphi$ are ranges of $\Delta y,\Delta\varphi$ space axes, $N_{\mathrm{av}}$ is the average amount of particles produced in analyzed multiplicity classes defined as
\begin{equation}
    N_{\mathrm{av}}= \langle {\frac{\mathrm{d} N_{\mathrm{ch}}}{\mathrm{d}y}} \rangle.
\end{equation}

\section{Analysis details}
The analysis is based on data from pp collisions at $\sqrt{s}$ = 13 TeV recorded by ALICE during 2016, 2017, and 2018 period. 
The angular correlation functions of pion, kaon, and proton pairs were analyzed for four multiplicity classes corresponding to 0--20\% (highest multiplicity), 20--40\%, 40--70\%, and 70--100\% (lowest multiplicity) of the total interaction cross section. The data sample consists of $\sim$ $1.4\times10^9$ minimum bias events. 
The analysis uses tracks reconstructed in the kinematic acceptance of $|y|$ $<$ 0.5, with 0.2 $<$ $p_{\mathrm{T}} < 2.5 \mathrm{GeV}/c$ for pions, and $0.5 < p_{\mathrm{T}} < 2.5 \mathrm{GeV}/c$ for kaons and protons. 
Particle identification is conducted on each track using TPC and TOF detectors \cite{detectorsdescription}. The identification is based on the $\rm{N_{\mathrm{\sigma}}}$ method, where $\sigma$ represents the standard deviation from the specific energy loss curves in TPC or the deviation of the time of flight from the anticipated arrival time of the particle in TOF. 
Based on the difference between the measured signal ``a" of a given track and the expected signal for pions, kaons, or protons in the TPC and TOF, particles with $N_{\mathrm{\sigma,PID}}^{\mathrm{a}} < 2$ were selected. Furthermore, for $p_{\mathrm{T}} > 0.5 GeV/c$, the values are obtained from the combined information of the TPC and TOF detectors, ${N_{\mathrm{\sigma,PID}}^{\mathrm{a}}}^{\mathrm{2}} = {N_{\mathrm{\sigma,TOF}}^{\mathrm{a}}}^{\mathrm{2}} + {N_{\mathrm{\sigma,TPC}}^{\mathrm{a}}}^{\mathrm{2}}$ and for $p_{\mathrm{T}} < 0.5 GeV/c$, information is obtained from the TPC only,  $N_{\mathrm{\sigma,PID}}^{\mathrm{a}} = N_{\mathrm{\sigma,TPC}}^{\mathrm{a}}$. The tracks for which the $N_{\mathrm{\sigma,PID}}^{\mathrm{a}} <$ 3 condition is fulfilled for more than one particle species hypothesis, are rejected.

\section{Results and discussions}
\begin{figure}[h!]
    \centering
    \includegraphics[width=4cm]{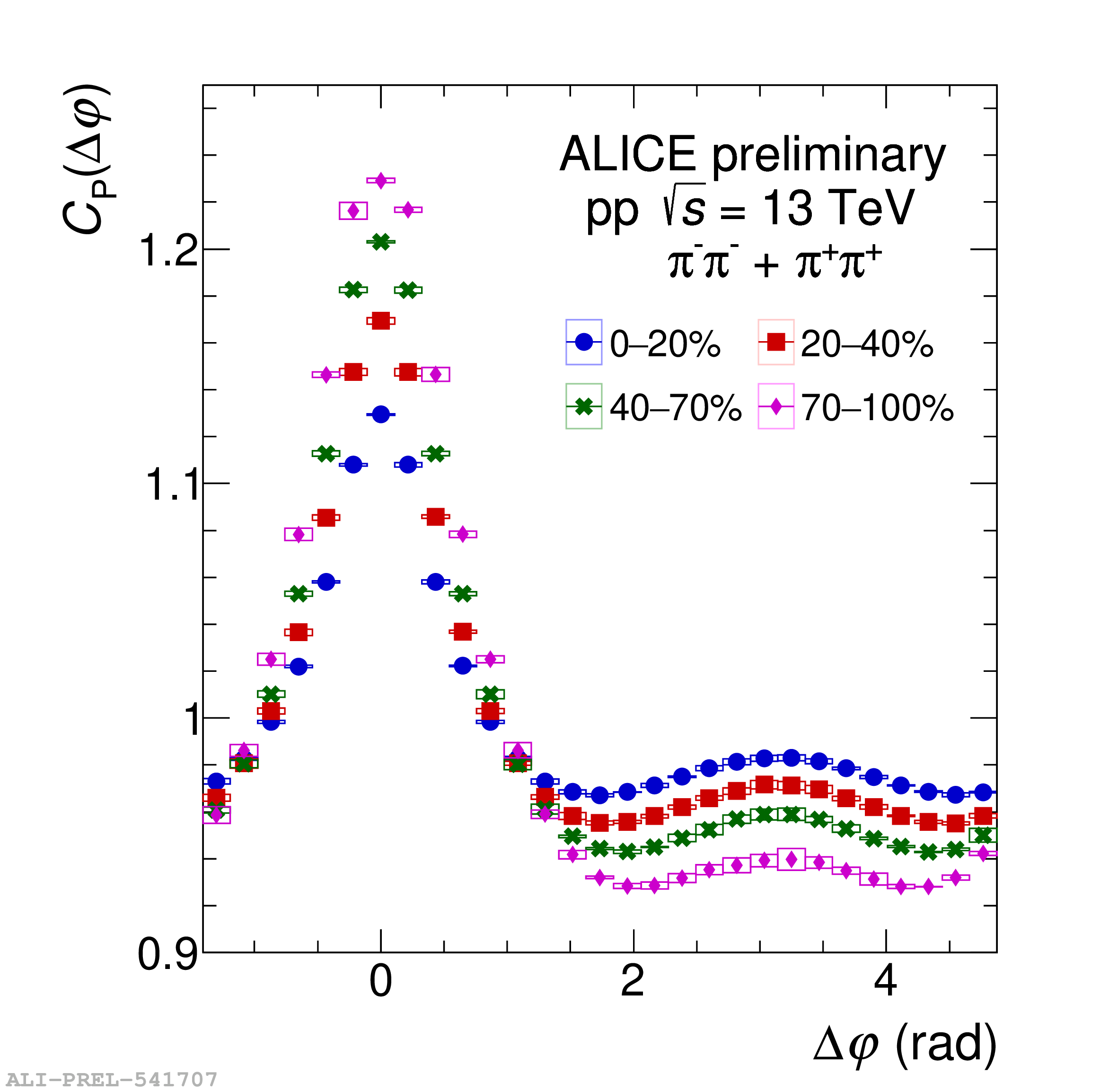}
    \includegraphics[width=4cm]{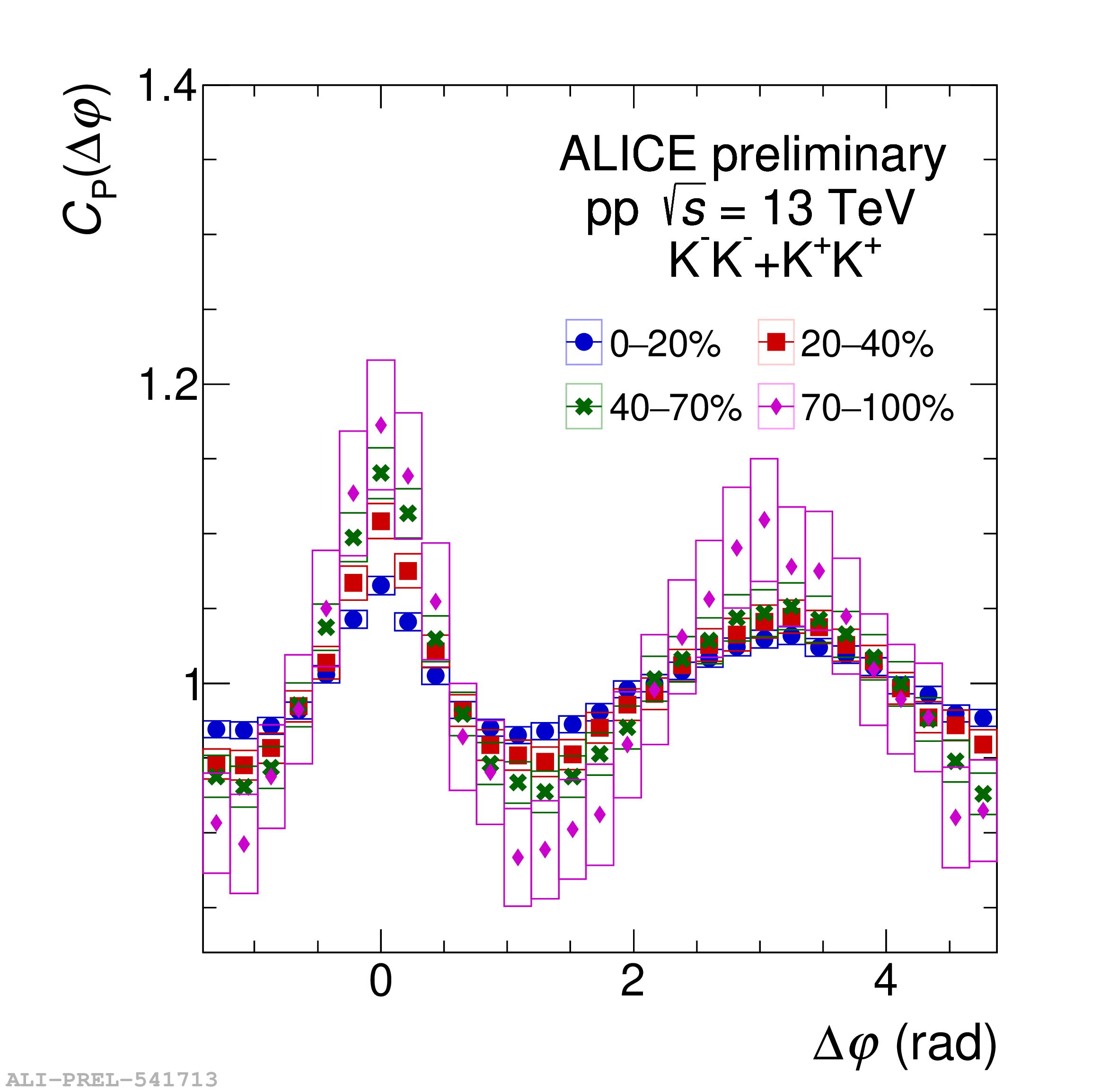}
    \includegraphics[width=4cm]{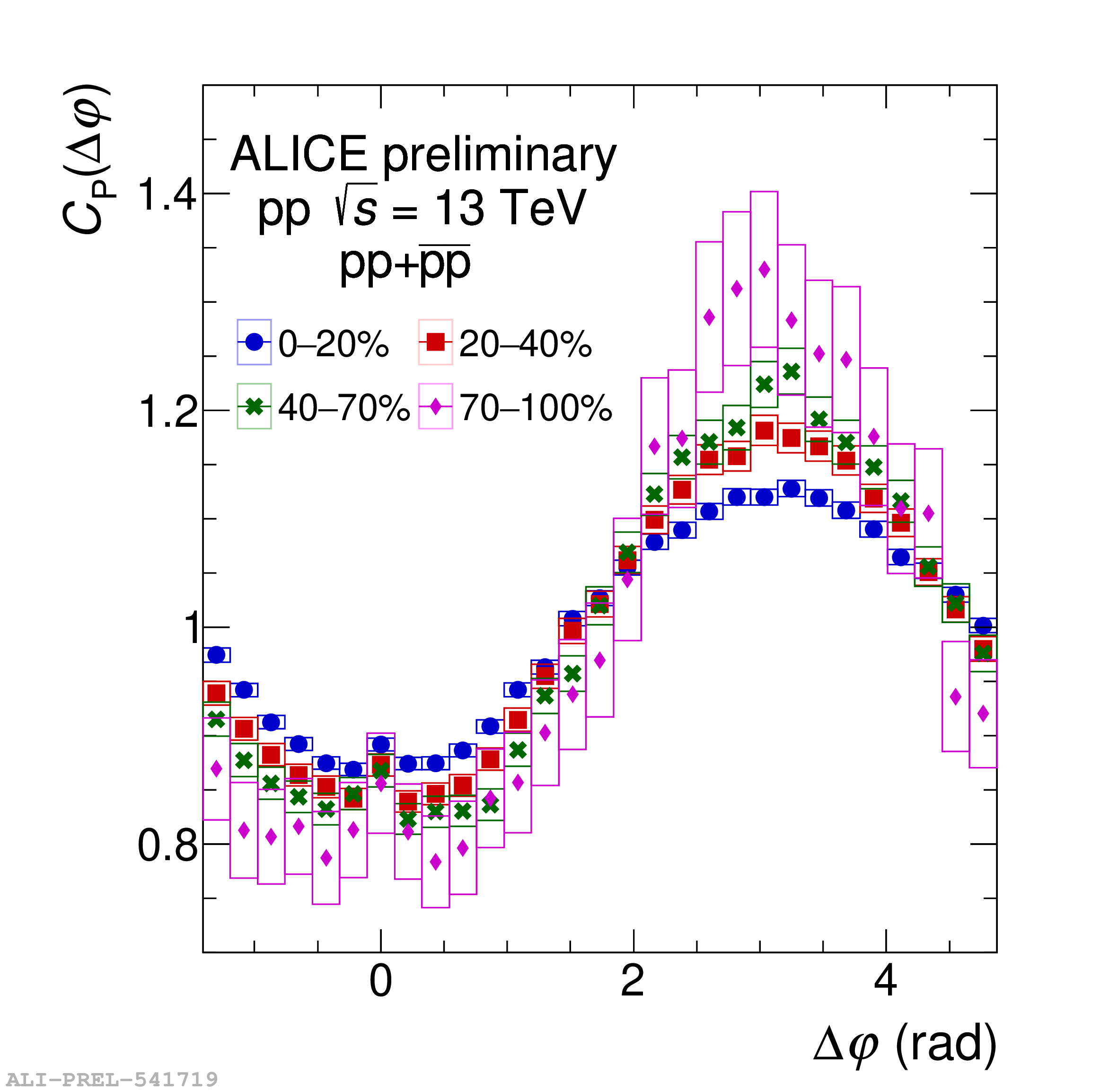}
    \includegraphics[width=4cm]{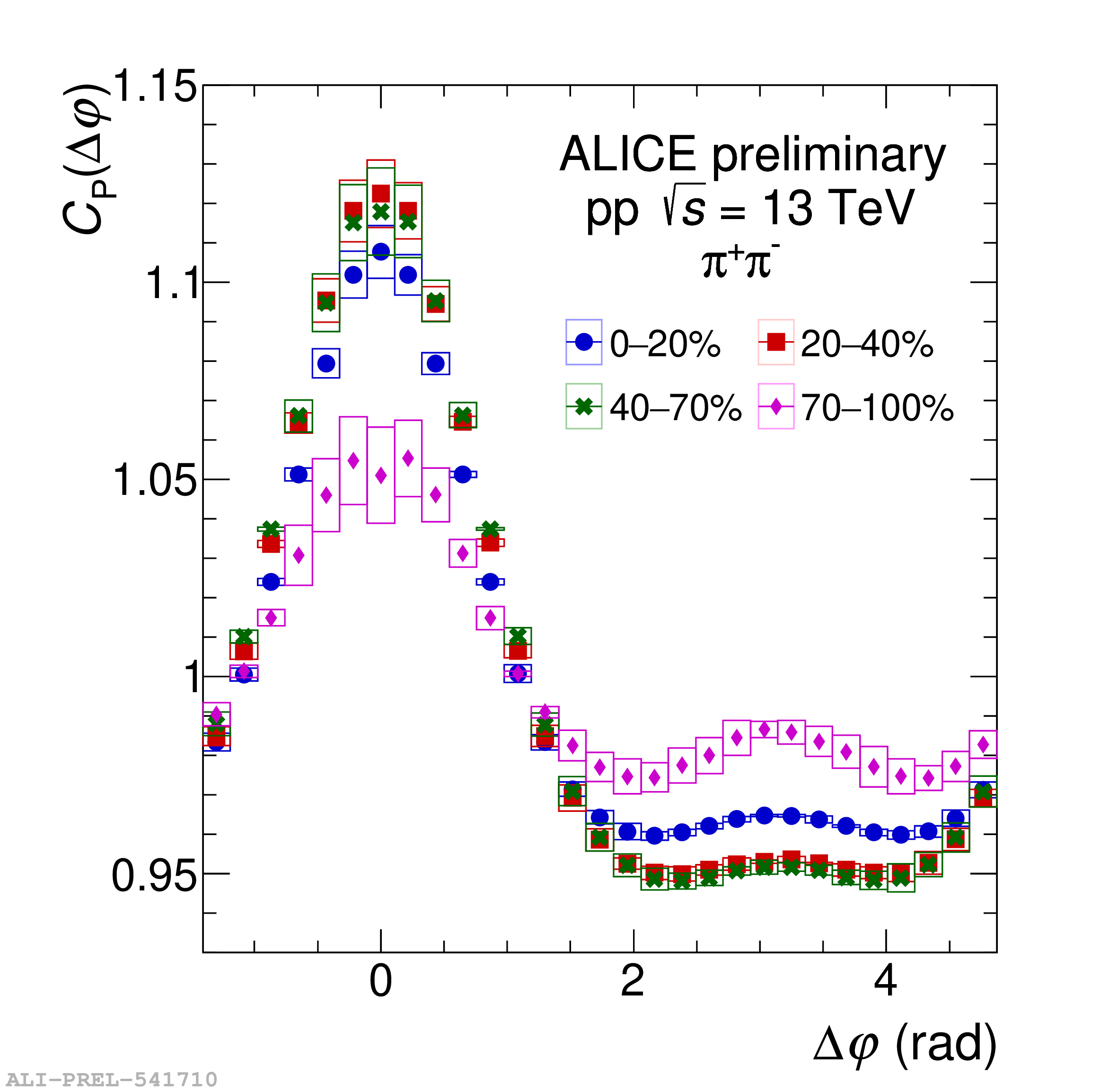}
    \includegraphics[width=4cm]{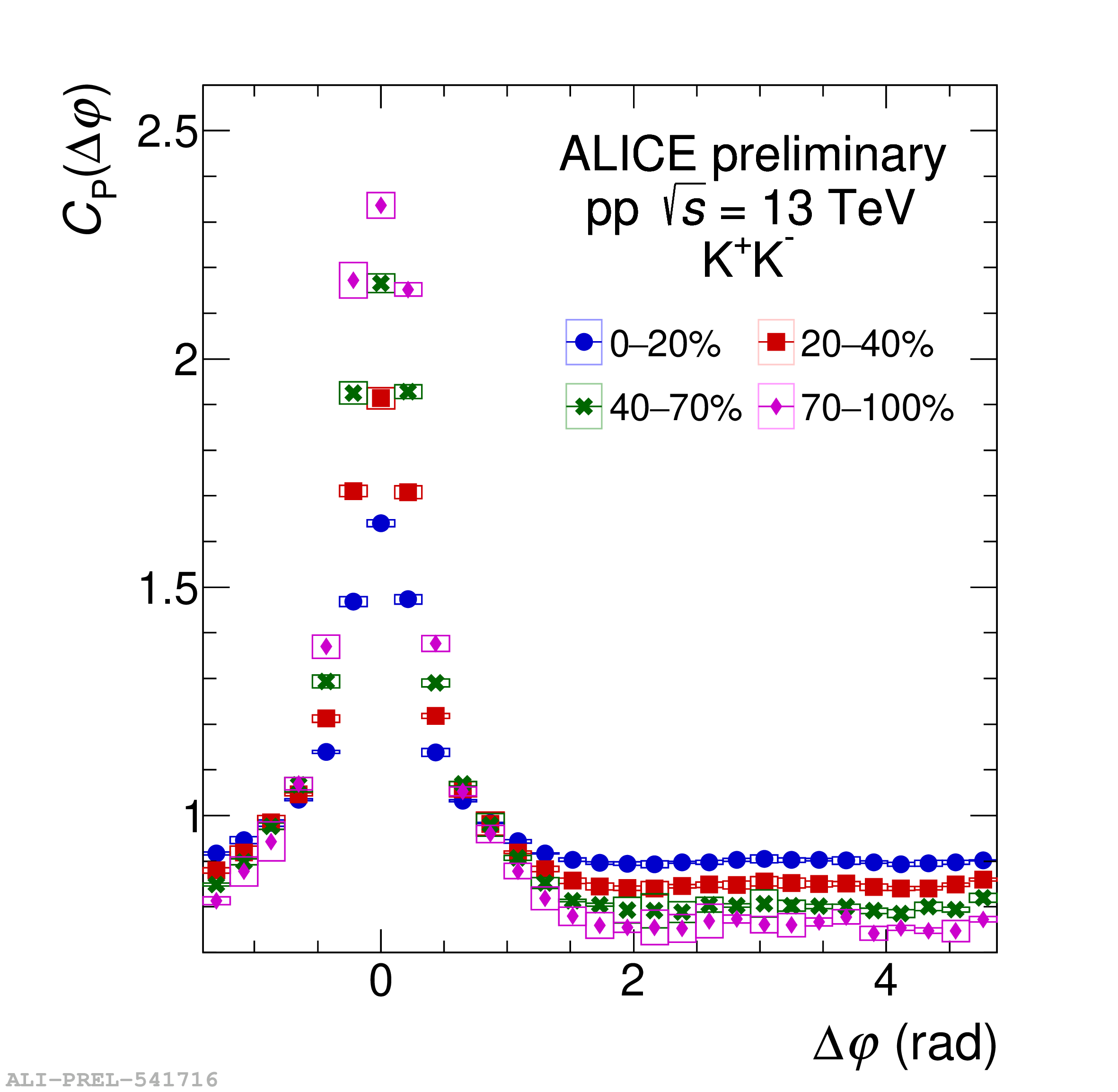}
    \includegraphics[width=4cm]{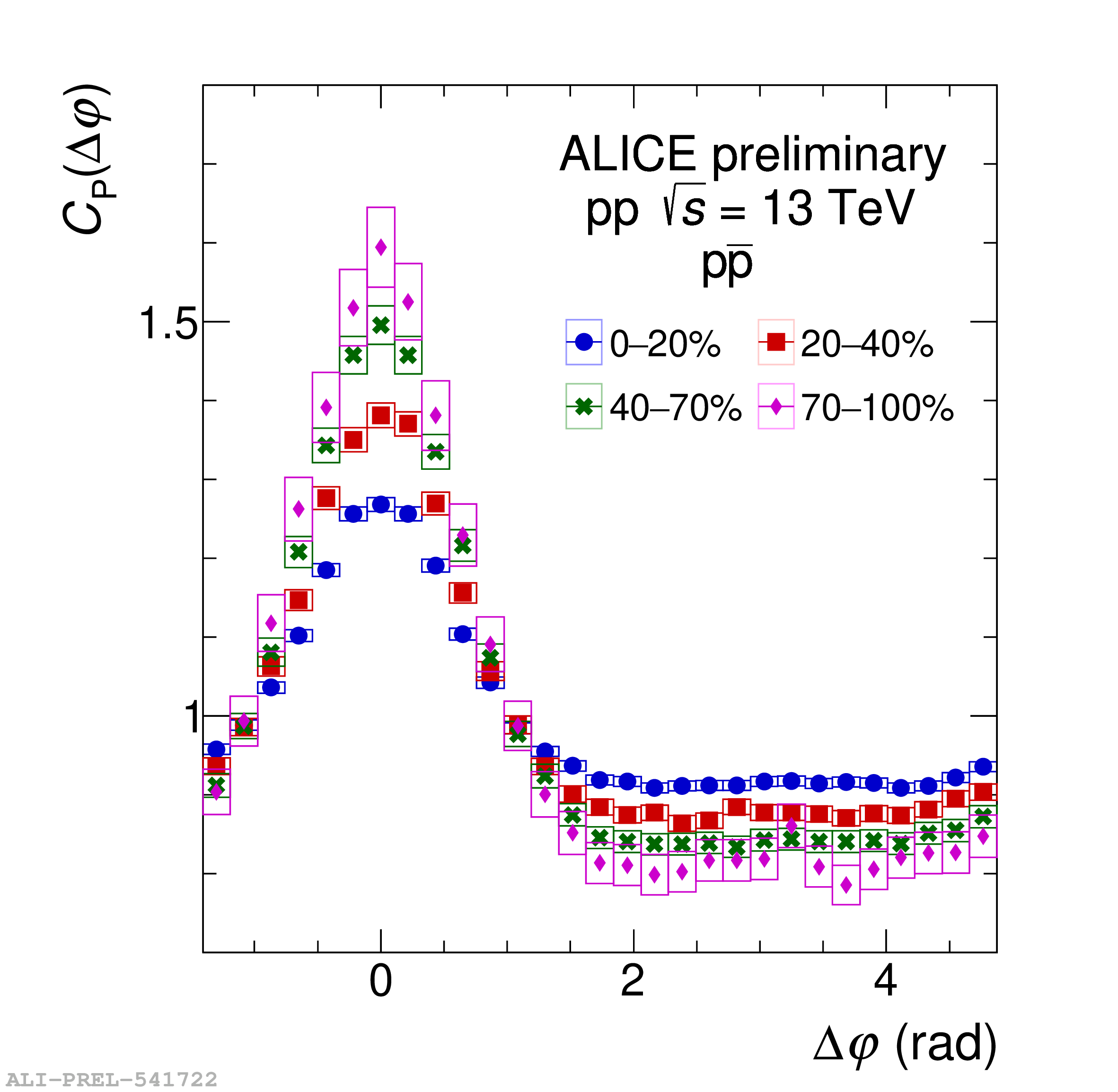}
    \caption{Projection of $\Delta y,\Delta\varphi$ correlation functions using the probability ratio definition in pp collisions at $\sqrt{s}$ = 13 TeV. Particles with like-signs are plotted on the first row, while those with unlike-signs are shown on the second row. For each particle, the four multiplicity classes are drawn on the same canvas.}
    \label{projection_probratio}
\end{figure}

\begin{figure}[h!]
    \centering
    \includegraphics[width=4cm]{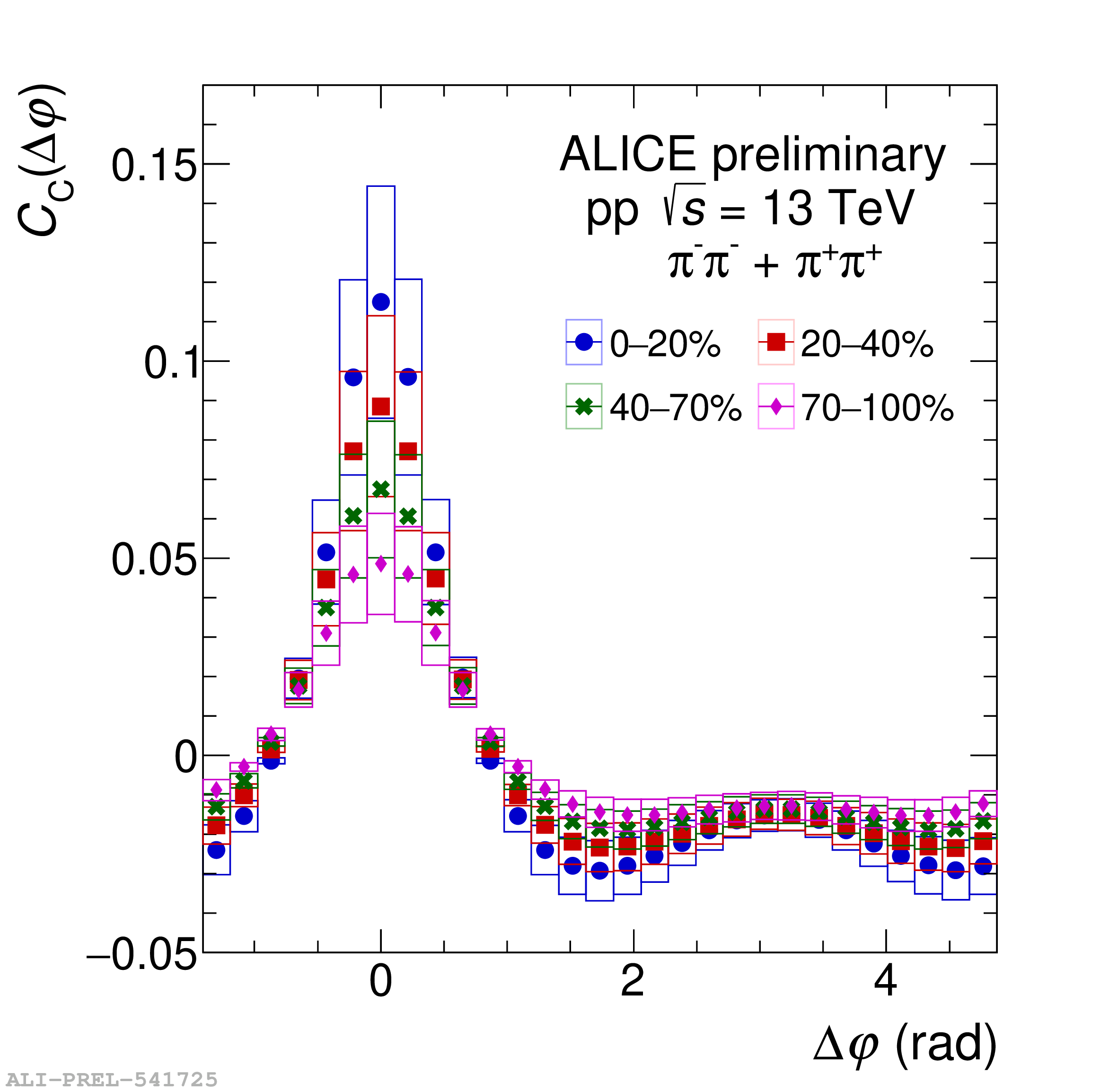}
    \includegraphics[width=4cm]{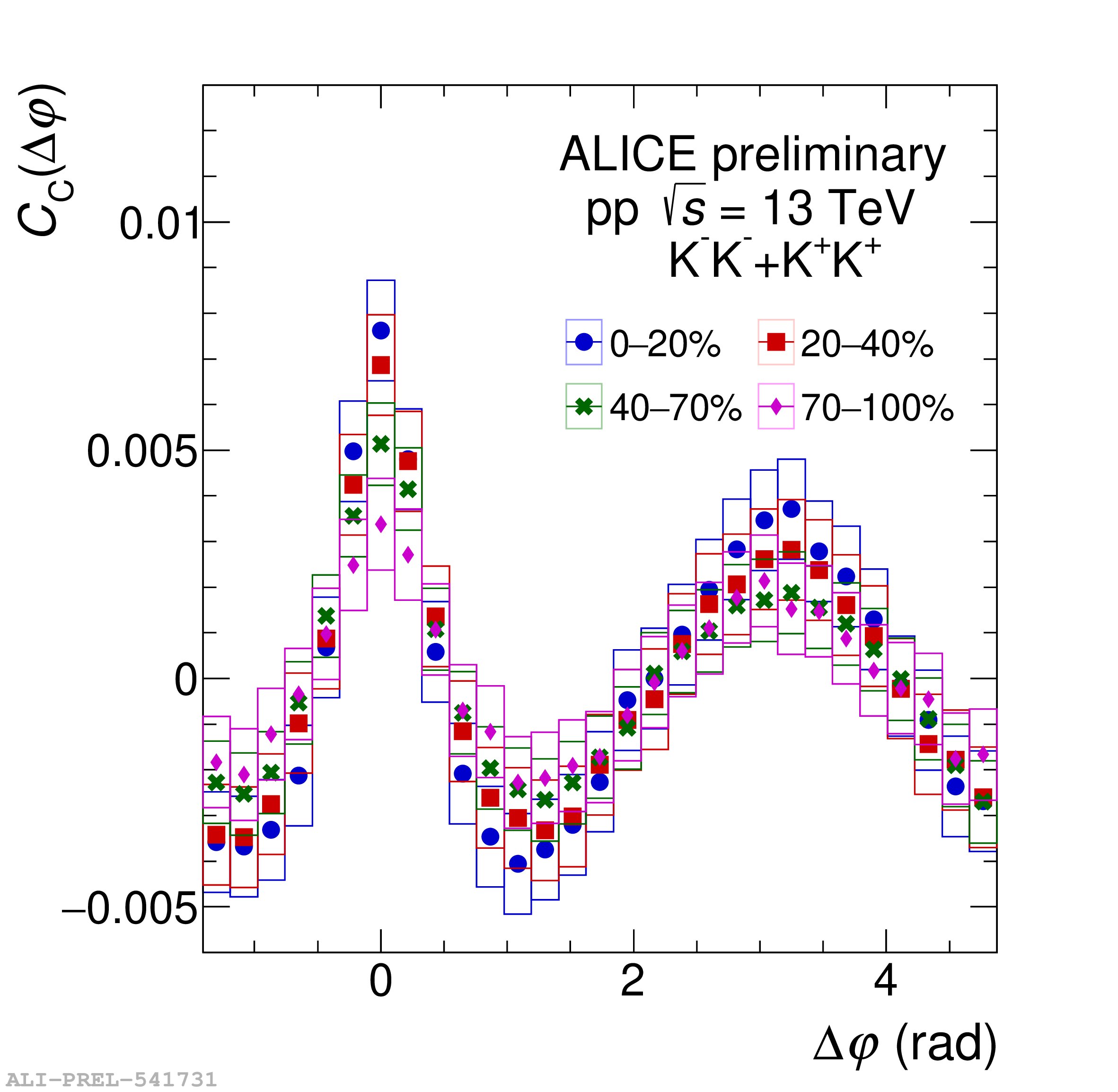}
    \includegraphics[width=4cm]{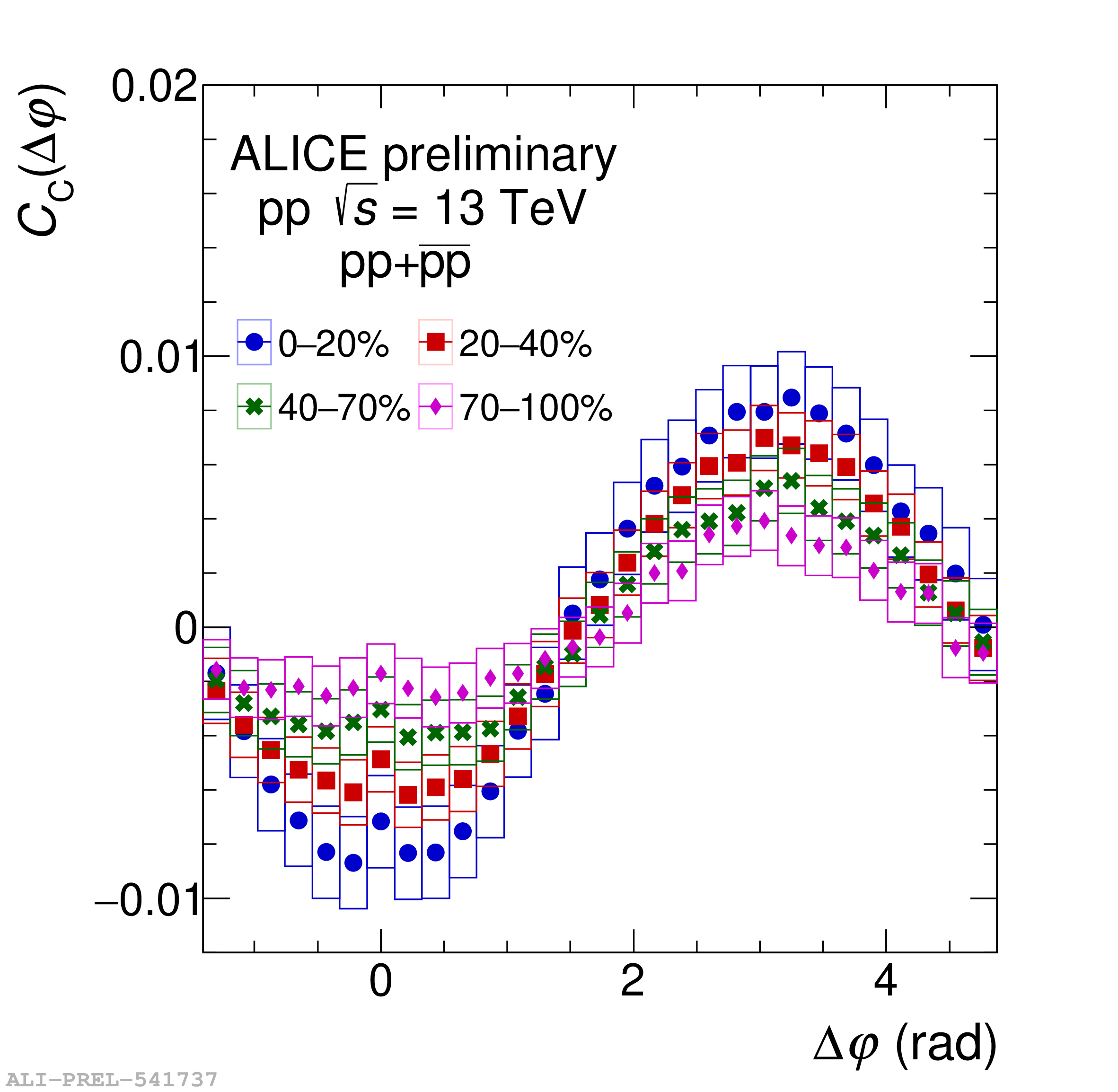}
    \includegraphics[width=4cm]{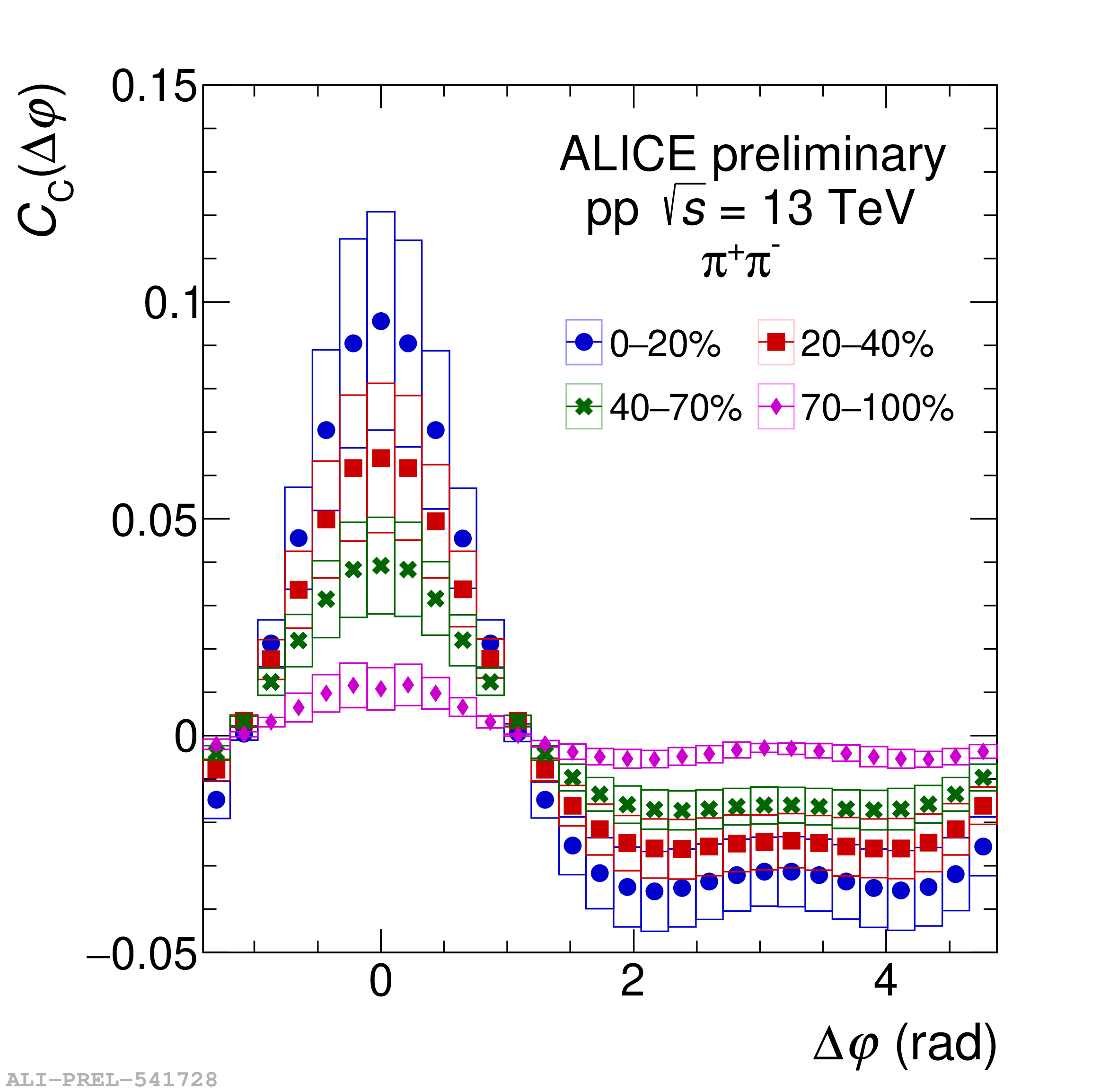}
    \includegraphics[width=4cm]{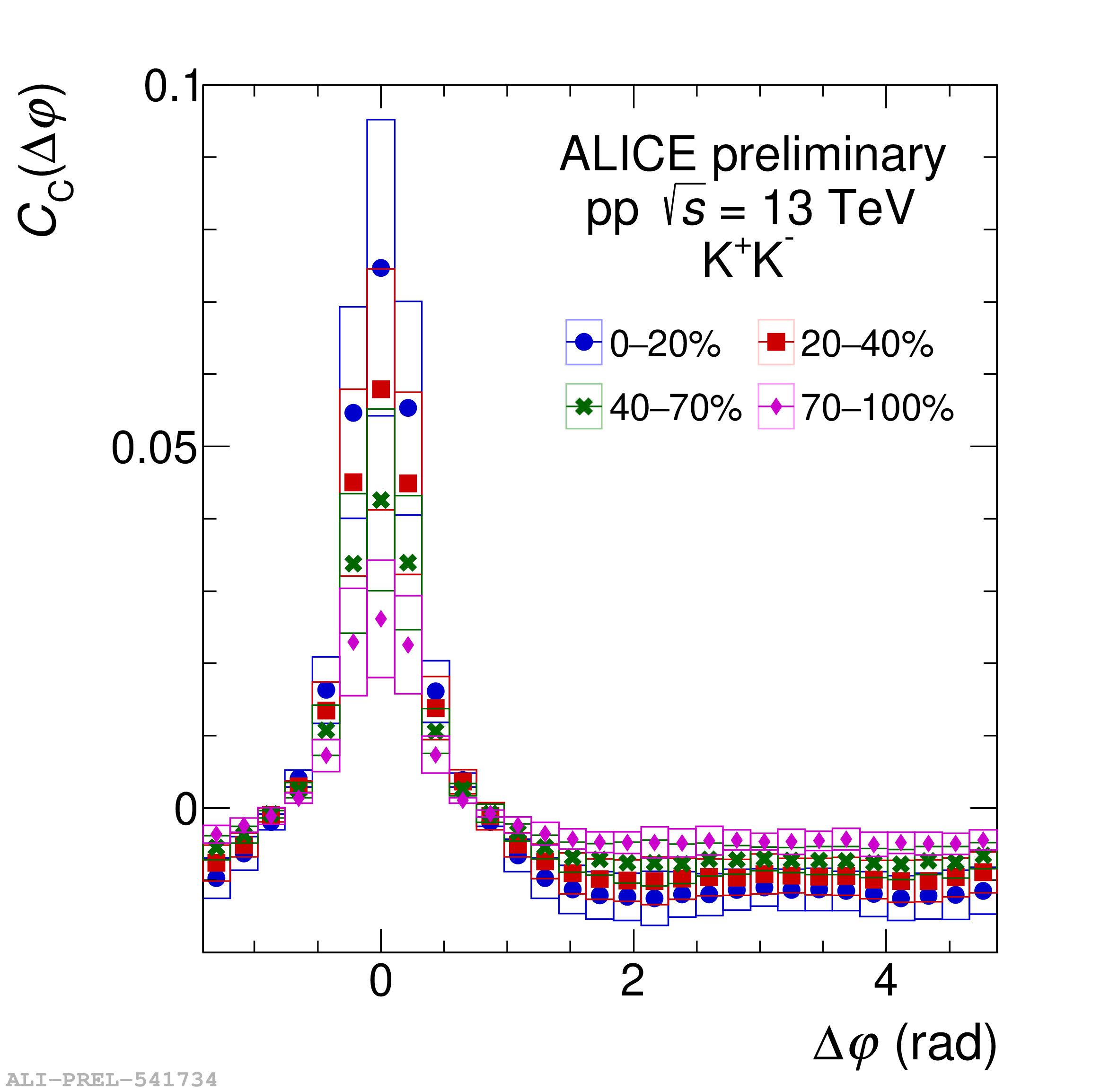}
    \includegraphics[width=4cm]{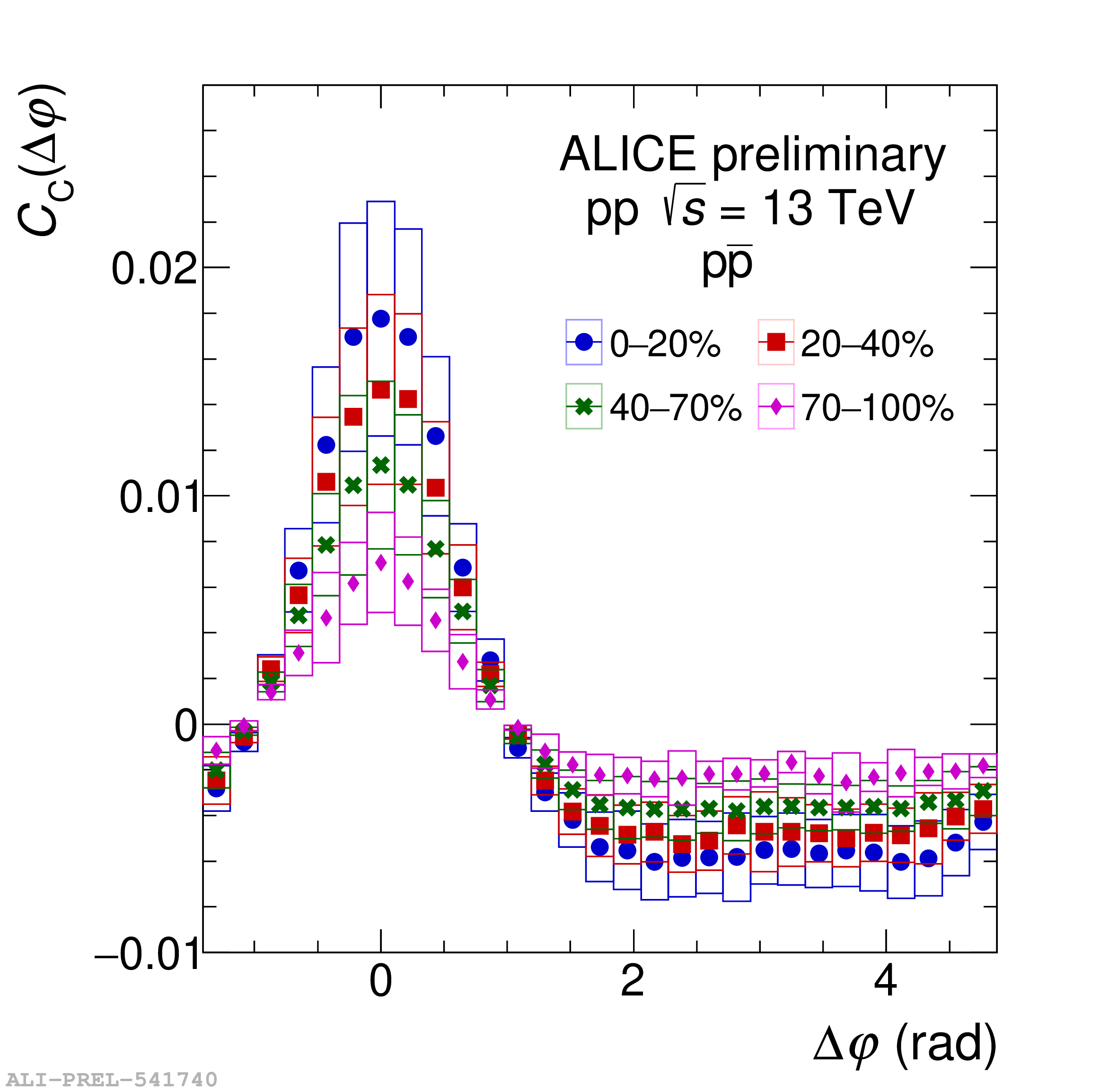}
    \caption{Projection of $\Delta y,\Delta\varphi$ correlation functions using the rescaled two-particle cumulant definition in pp collisions at $\sqrt{s}$ = 13 TeV. Particles with like-signs are plotted on the first row, while those with unlike-signs are shown on the second row. For each particle, the four multiplicity classes are drawn on the same canvas.}
    \label{projection_rescaled}
\end{figure}

The correlation functions for pions, kaons, and protons pairs measured using probability ratio definition and rescaled two-particle cumulant in the four multiplicity classes are displayed in Figs. \ref{projection_probratio} and \ref{projection_rescaled}, respectively. The results of both definitions show a visible peak in the near-side region for pion and kaon pairs with same-sign correlations, attributed to various effects including fragmentation of hard-scattered partons (also known as minijets) and Bose--Einstein correlations. Parton scattering processes involving high mass transfers on short timescales result in the production of collimated quarks and gluons. These partons hadronize and produce collimated streams of particles known as jets. At low momenta, these particles, called minijets, are produced at a small difference in azimuthal and polar angles, approximately equal to ($\Delta y,\Delta\varphi$) $\sim$ (0,0), contributing to the near-side peak. In addition, the peak is larger in the case of bosons than in the case of fermions, because there is a greater probability of finding two bosons in the same quantum state, whereas the Pauli exclusion principle prohibits the existence of two similar fermions in the same state. Therefore, the Bose--Einstein quantum statistics has an impact on the structure of the near-side peak, exclusively for pairs of same-sign particles. 
The baryon--baryon and antibaryon--antibaryon distributions for pairs of identical protons show a qualitatively different effect. They show a wide anticorrelation on the near-side instead of a peak, together with a ridge on the away-side, as already observed in pp collisions at $\sqrt{s}$ = 7 TeV \cite{Adam_2017}. In contrast, particle--antiparticle correlations for mesons and baryons show a minijet structure on the near-side and a weak structure on the away-side. Bose--Einstein and Fermi--Dirac effects do not occur in pairs of nonidentical particles, but resonances significantly affect the correlation function. The correlation between mesons and baryons is similar in quality. The difference lies in the amplitude and width of the near-side peak, which in fact is higher for kaons than for protons and pions. The shape and strength of the correlation functions, which have a distinct near-side peak, show that they can be influenced mainly by considerable contributions from minijets.\\
\indent
Figure \ref{projection_probratio} shows the correlation functions defined as probability ratio and labeled as $C_{\mathrm{P}}(\Delta y,\Delta\varphi)$ \cite{probability}, while Fig. \ref{projection_rescaled}  shows the rescaled two-particle cumulant functions labeled as $C_{\mathrm{C}} (\Delta y,\Delta\varphi)$ \cite{rescaled}. Using the definition of the probability ratio, the correlation functions increase toward lower multiplicity classes, due to the scaling of the multiplicity by 1/N. On the contrary, the rescaled two-particle cumulant shows an opposite trend, indicating an increase in the correlation function with higher multiplicity.

\begin{figure}
    \centering
    \includegraphics[width=11cm]{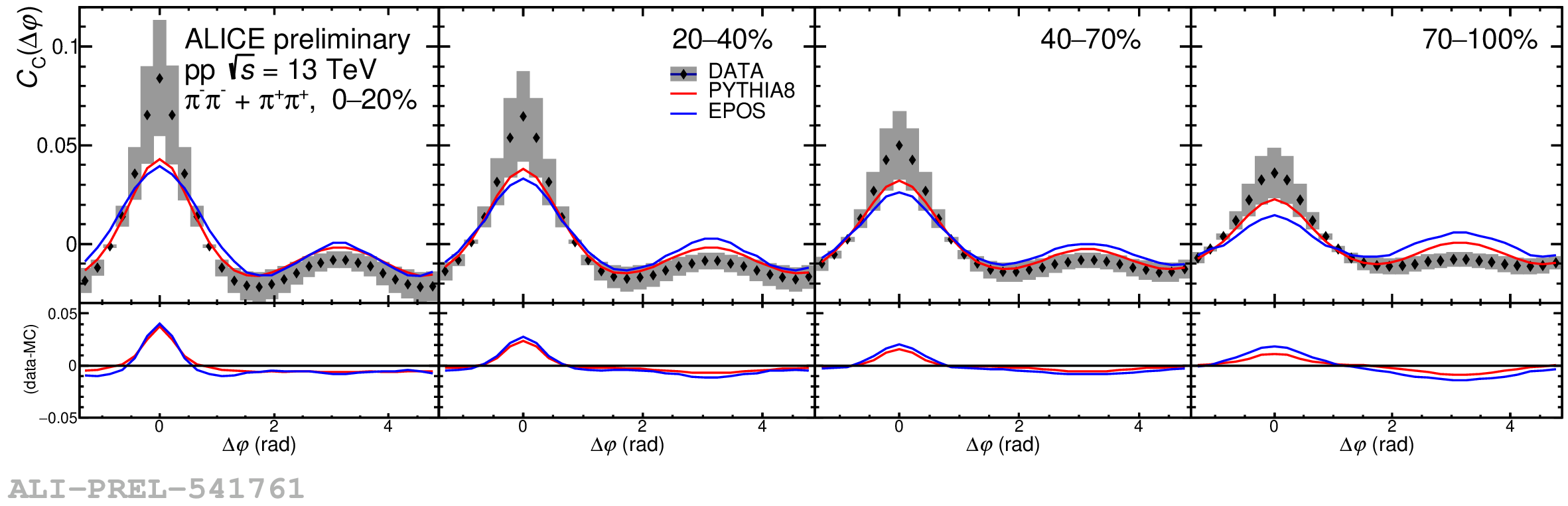}
    \includegraphics[width=11cm]{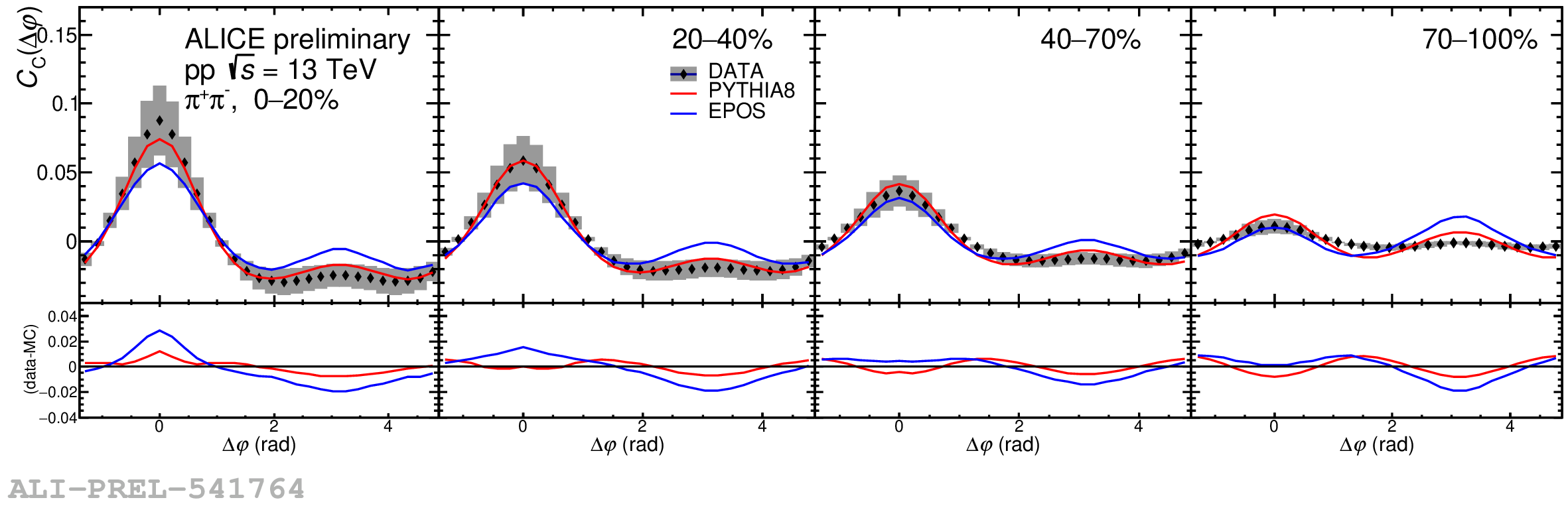}
    \includegraphics[width=11cm]{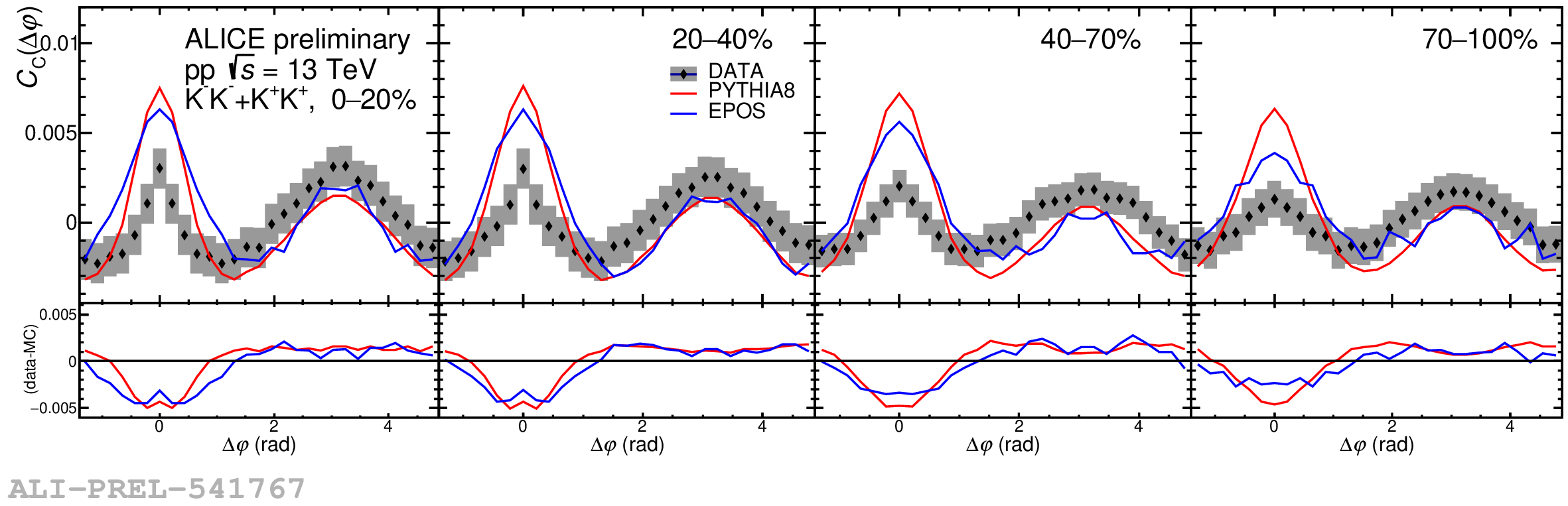}
    \includegraphics[width=11cm]{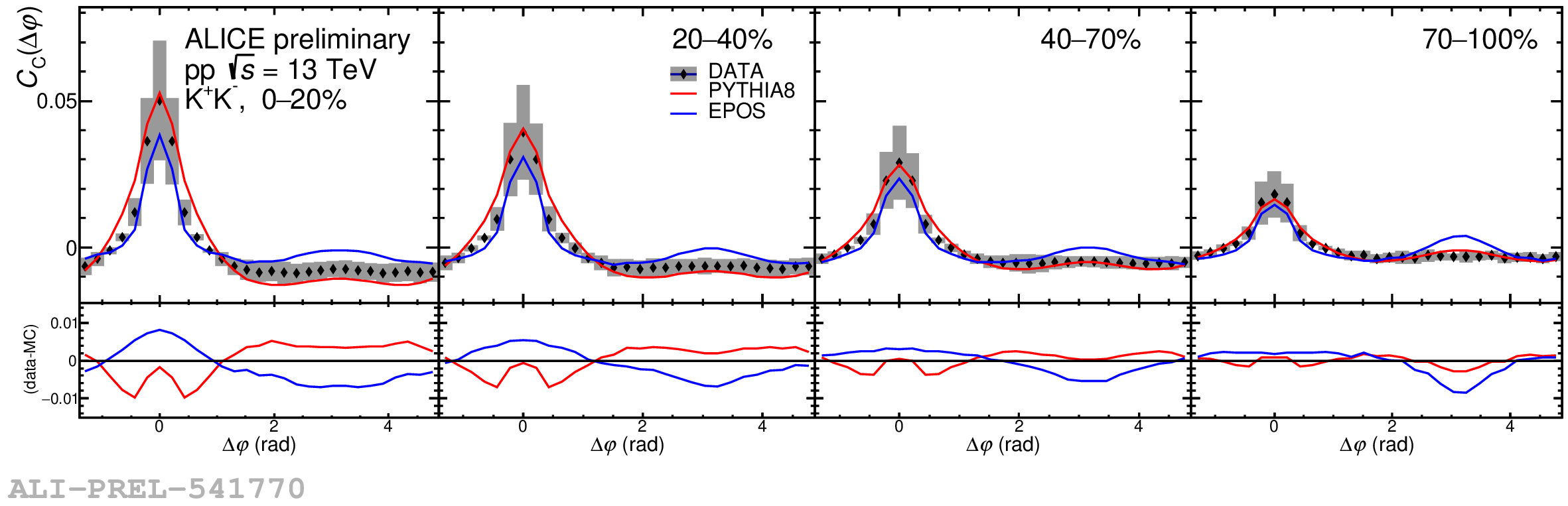}
    \includegraphics[width=11cm]{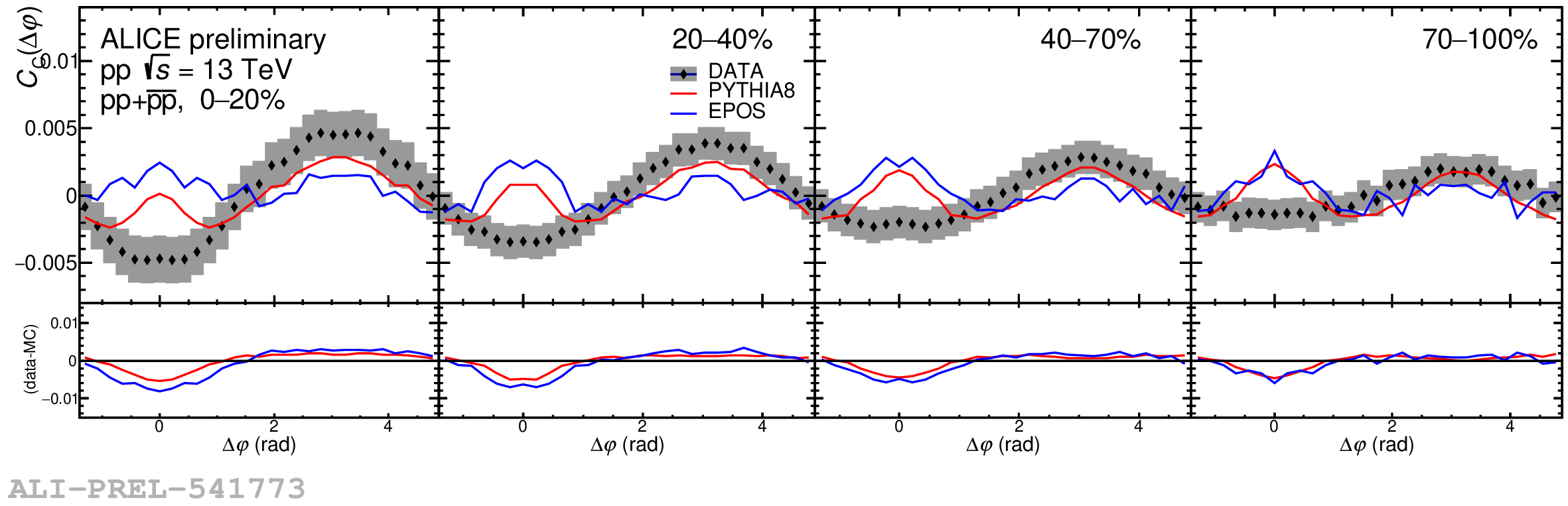}
    \includegraphics[width=11cm]{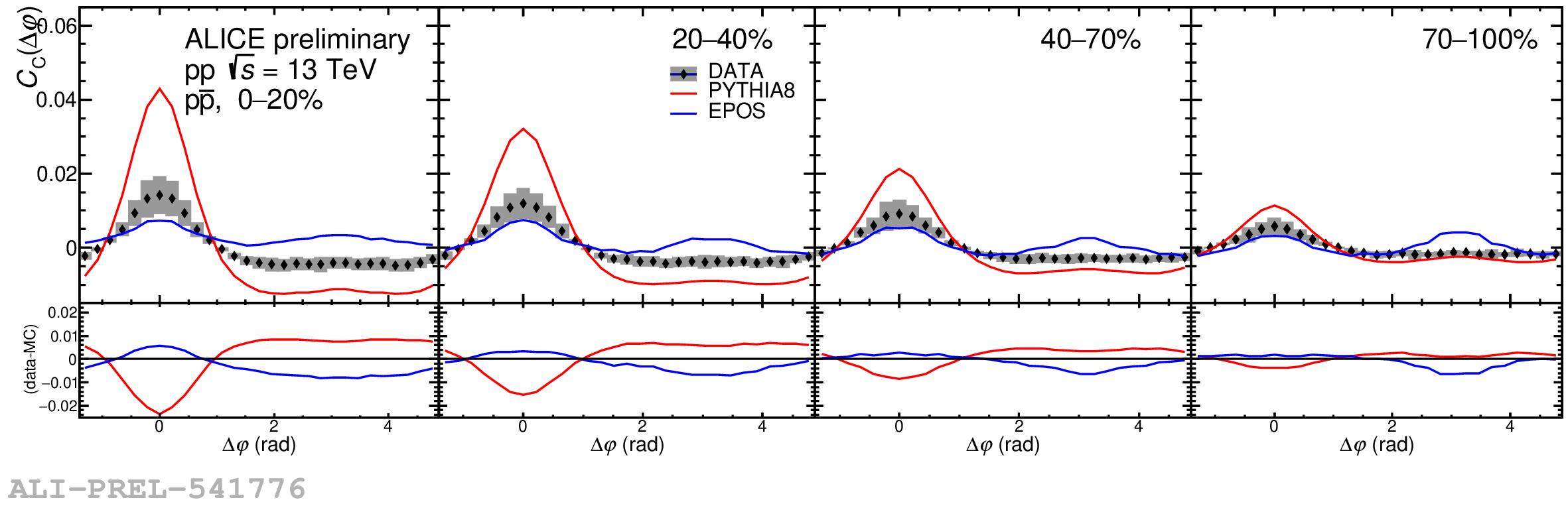}
    \caption{Comparison of $\Delta y,\Delta\varphi$ correlation functions using the rescaled two-particle cumulant definition with MC models in pp collisions at $\sqrt{s}$ = 13 TeV. The comparison is computed for pions, kaons, and protons pairs with like and unlike-sign and for the multiplicity classes 0--20\%, 20--40\%, 40--70\%, and 70--100\%. The bottom panels show relative differences between each model and the data.}
    \label{comparisonwithmodels}
\end{figure}

In order to gain a deeper understanding based on the observations made, it is important to compare the results derived from the definition of the rescaled two-particle cumulant with the theoretical Monte Carlo models, PYTHIA8 \cite{pythia8} and EPOS \cite{epos} both at $\sqrt{s}$ = 13 TeV, which are widely used for simulations of high-energy collisions. The Monte Carlo model calculations shown in Fig. \ref{comparisonwithmodels} reproduce the experimental results for mesons reasonably well. However, the models fail to reproduce the correlations for baryons (both particle-particle and particle-antiparticle pairs). For protons, no anticorrelation is observed in any of the models studied.

\section{Conclusions}
The research conducted in this work involved the analysis of correlation functions in $\Delta y,\Delta\varphi$ space using alternative definitions, namely the probability ratio and the rescaled two-particle cumulant. The results reveal that the correlation functions obtained by the rescaled two-particle cumulant  in different multiplicity classes are displayed with an increasing tendency as the multiplicity classes increase. Furthermore, the implementation of the following definition allows us to explore in depth the physical phenomena that contribute to the overall shape of the correlation function, eliminating the trivial scaling of 1/N. Comparing our results with those obtained from the PYTHIA8 and EPOS models, a fair experimental agreement with mesons is observed, while the models fail to reproduce baryon correlations. In particular, the baryon--baryon (antibaryon--antibaryon) pairs show a considerable depletion called anticorrelation that is not qualitatively reproduced by the models. Whereas, the correlations simulated by the baryon--antibaryon pairs are qualitatively similar to those observed in the experimental data, but much stronger in the models. This study will be extended by comparing the pp results with those from p--Pb and Pb--Pb collisions.

\section{Acknwledgments}
This work was supported by the Polish National Science Centre under agreements \textcolor{blue}{UMO-2021/43/D/ST2/02214} and \textcolor{blue}{UMO-2022/45/B/ST2/02029}, by the Polish Ministry for Education and Science under agreements no. \textcolor{blue}{2022/WK/01} and \textcolor{blue}{5236/CERN/2022/0}, as well as by the \textcolor{blue}{IDUB-POB-FWEiTE-3} project granted by Warsaw University of Technology under the program Excellence Initiative: Research University (ID-UB).
\renewcommand{\refname}{} 
\renewcommand\refname{}

\end{document}